\documentclass[
 reprint,
 amsmath,amssymb,
 aps,
]{revtex4-1}

\usepackage{graphicx}
\usepackage{dcolumn}
\usepackage{bm}
\usepackage{textgreek}

\begin{document}

\preprint{APS/123-QED}

\title{
Dynamic swarms regulate the morphology and distribution of soft membrane domains
}

\author{Aakanksha Gubbala}
\email{A.G. and D.P.A. contributed equally to this work}
\author{Daniel P. Arnold$^*$}
\author{Anika Jena}
\author{Stephanie Anujarerat}

 \author{Sho C. Takatori}
 \email{stakatori@ucsb.edu}
\affiliation{
 Department of Chemical Engineering, University of California, Santa Barbara, Santa Barbara, CA 93106
}

\begin{abstract}
We study the dynamic structure of lipid domain inclusions embedded within a phase-separated reconstituted lipid bilayer in contact with a swarming flow of gliding filamentous actin.
Passive circular domains transition into highly-deformed morphologies that continuously elongate, rotate, and pinch off into smaller fragments, leading to a dynamic steady state with $\approx 23 \times$ speed up in the relaxation of the intermediate scattering function compared to passive membrane domains driven by purely thermal forces.
To corroborate experimental results, we develop a phase-field model of the lipid domains with two-way coupling to the Toner-Tu equations. 
We report phase domains that become entrained in the chaotic eddy patterns, with oscillating waves of domains that correlate with the dominant wavelengths of the Toner-Tu flow fields.

\end{abstract}


\maketitle

\section{Introduction}

Large-scale reorganization of the actin cytoskeleton can generate collective flows against the plasma membrane in both living cells \cite{Hueschen2022,Zonies2010} and in-vitro model membranes \cite{ArnoldGubbala2023,Koster2016,Liebe2023}.
Lipid membranes form a quasi-2-dimensional (2D) in-plane fluid that can laterally phase-separate into distinct domains of both proteins and lipids \cite{Veatch2003,Veatch2008,Lee2023,Hsu2023a}, spatially patterning membrane mechanical and chemical properties \cite{Yuan2021,Zeno2016,Pettmann2018,Arnold2023}.
In mammalian cells, clusters of proteins and lipids on the cell surface assemble and disassemble in response to chemical signals \cite{Lingwood2010,Levental2020}, with some domains, like the T-cell synapse, reaching several microns in size \cite{Nye2008}.

Actomyosin networks have been shown to interact with phase-separated domains on lipid bilayers in-vitro \cite{Honigmann2014,Arnold2023a}, and the actin cytoskeleton is known to be involved in the growth of membrane domains like the B-cell synapse \cite{Viola2007}.
Previous in-vitro measurements on lipid membranes with adsorbed actin networks have focused on the effects of a single contractile pulse on the static structure of domains \cite{Koster2016,Vogel2017}.
Recently we found that membrane domains in contact with a rapidly-contracting actomyosin network exhibit a 2$\times$ acceleration in domain growth compared to passive thermal coarsening \cite{ArnoldGubbala2023}.
In this article, we consider gliding actin filaments, propelled along a solid substrate by stationary motors, in contact with a multiphase lipid membrane.
The shift from a contractile 2D network to swarms of aligned filaments gives rise to a dynamic steady state characterized by mixing of lipid domains, instead of accelerated coarsening.

Driving liquid-liquid phase separated (LLPS) systems out of equilibrium using active microtubule/kinesin flows provides a means of altering droplet morphology, size, and wetting properties \cite{Adkins2022, Tayar2023, Caballero2022}, and has received significant recent interest in 3-dimensional (3D) systems.
In contrast with the $\approx$10-100 \textmu m 3D droplets studied in these works, our article considers 2D lipid domains of diameter $\sim$1-10 \textmu m in a membrane of thickness $\approx$4 nm. 
Moreover, Tayar et al. required chemical cross-links between DNA nanostar droplets and kinesin motors to elicit interactions between the droplets and active microtubules in the continuous phase \cite{Tayar2023}.
In our two-dimensional lipid membrane, the time scale of stress transmission from the active filaments to the droplets is fast relative to that of stress dissipation.
Thus, we achieve significant stress transmission from the active filaments via the actin-coupled Ld phase, without the need for any cross-linkers between actin and Lo.

Prior authors have used nematohydrodynamic theory, coupled with Cahn-Hilliard phase-field modeling to describe the effect of active nematic flows on phase-separated materials and interfaces \cite{Caballero2022,Xu2023}.
We instead use the Toner-Tu flocking model \cite{Toner1995,Toner1998}, which has been used to describe swarming bacteria \cite{wensink_meso-scale_2012}, chemotaxis \cite{Miller2024}, and actin flows along the cell surface \cite{Hueschen2022}.
Hueschen et al. recently presented polar flocking flows of actin along the surface of \textit{Toxoplasma gondii}, driven by stationary myosin motors within the cell \cite{Hueschen2022}.
Using the Toner-Tu equation to model actin flocks along a curved surface, these authors found that \emph{T. gondii} motion switched from a bidirectional cyclosis mode to a unidirectional helical mode when a reaction term was added to permit actin polymerization and depolymerization at the poles of the cell.
These results closely matched experimental data showing that actin recirculation along the cell surface controlled the mode of cellular motion, confirming that the Toner-Tu model effectively describes actin flow behavior on surfaces.

Our actin flows along a solid substrate organize into polar flocks qualitatively similar to those studied by Hueschen et al. \cite{Hueschen2022}.
Thus, instead of a nematohydrodynamic approach, we use the Toner-Tu equations to describe the polar flocks, in contact with a multiphase 2D fluid.
By coupling the Toner-Tu equations with the Cahn-Hilliard phase field model, our goal is to study the dynamic evolution of the domain morphology and distribution subjected to chaotic active flow fields.

\section{Methods and Materials}

\subsection{Experiments}
\begin{figure}
    \centering
    \includegraphics[width=\linewidth]{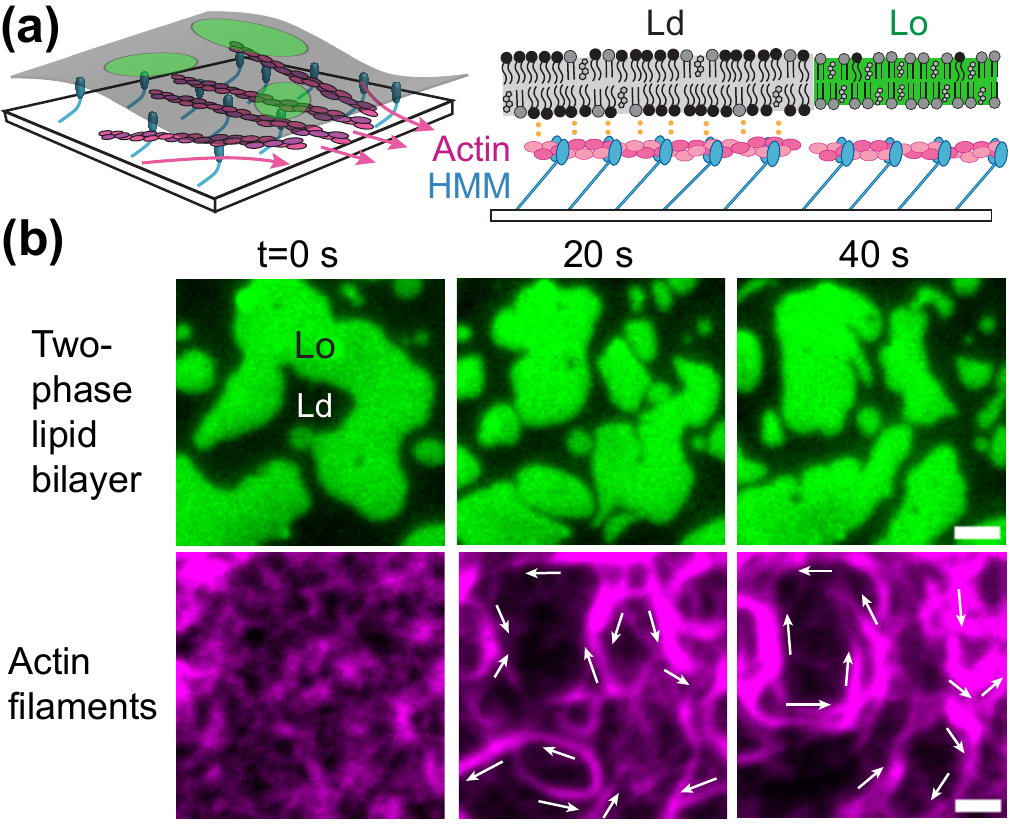}
    \caption{
    Fluid membrane domains driven by swarming filamentous actin (F-actin) transition into highly-deformed morphologies that continuously elongate, rotate, and pinch off into smaller fragments.
    (a) Heavy meromyosin (HMM, blue) shuttles nematically-aligned F-actin (magenta) along a coverslip surface, consistent with the actin gliding assay \cite{Kron1986,Schaller2010}.
    A planar phase-separated lipid bilayer containing liquid-ordered (Lo) domains (green) is dispersed in a liquid-disordered (Ld) continuous phase (gray).  
    The gliding actin is electrostatically coupled to the Ld phase via charged lipids.
    (b) Fluorescence micrographs show the motion of Lo domains (top) to swarming actin filaments (bottom).
    When HMM becomes active at time $t$=0, actin is shuttled across the coverslip surface, forming flocks.
    These flocks generate in-plane flows in the membrane and cause continuous deformation, fission, and fusion of Lo domains.
    White arrows indicate direction of actin flows.
    Scale bar is 2 \textmu m.
    }
    \label{fig:overview}
\end{figure}

Heavy meromyosin (HMM) is adsorbed to glass coverslips treated with a hydrophobic coating of trimethylchlorosilane (see Supplemental Material for detailed methods).
Filamentous actin (F-actin) is bound to the S1 propulsive domain of HMM at actin densities above the flocking density, as in previous gliding assays \cite{Kron1986,Schaller2010}.
Giant unilamellar vesicles (GUVs) are formed with 45\% dioleoylphosphatidylcholine (DOPC), 35\% dipalmitoylphosphatidylcholine (DPPC), 15\% cholesterol, and 5\% dioleoyl-3-trimethylammonium propane (DOTAP) using established electroformation methods \cite{Angelova1986}.
At room temperature, these lipid bilayers phase-separate into liquid-ordered (Lo) domains embedded within a liquid-disordered (Ld) continuous phase.
GUVs rupture on top of actin and HMM at 50\textdegree C, driven by electrostatic attractions between the positively charged, Ld-favoring DOTAP and the negatively charged actin and glass substrate [Fig.~\ref{fig:overview}(a)].

Upon cooling to room temperature, Lo domains condense and coarsen to adopt non-circular morphologies, conforming with the underlying actin layer [Fig.~\ref{fig:overview}(b), left panels].
ATP is introduced at time $t$=0~s, triggering HMM to begin shuttling F-actin across the surface.
Gliding F-actin align with one another, forming dynamic swarms that assemble and dissolve over time [Fig.~\ref{fig:overview}(b), bottom row, Supplemental Video~S1].
The Lo domains rapidly deform, translate, divide, and fuse in response to the collective actin flows, losing any qualitative resemblance to the original Lo domain microstructure within 20~s [Fig.~\ref{fig:overview}(b), top row].

\subsection{Theory}

To gain analytical insight into how flocking flows affect lipid domains,
we model the lipid phase field by evolving a concentration order parameter $\phi(\mathbf{x}, t)$ using the Cahn-Hilliard equation:
\begin{equation}\label{eq:CahnHilliard}
\frac{\partial \phi}{\partial t} + \mathbf{v}\cdot\nabla\phi = M\nabla^2\mu \ ,
\end{equation}
with lipid mobility $M$, chemical potential $\mu = \delta f/\delta \phi - \kappa \nabla^2\phi$, interfacial tension parameter $\kappa$, and bulk free energy $f[\phi(\mathbf{x}, t)] = \phi^4/4 - \phi^2/2$ modeled as a double well potential where the concentration of pure phases is $\pm 1$.

The lipids are convected by a velocity $\mathbf{v}(\mathbf{x}, t)$, which is given by the Toner-Tu model for flocking flows \cite{Toner1995,dunkel_minimal_2013,wensink_meso-scale_2012}:
\begin{align}\label{eq:TonerTu}
\frac{\partial \mathbf{v}}{\partial t} + (1 - S)\mathbf{v}\cdot\nabla\mathbf{v} =& -\frac{S}{2}\nabla |\mathbf{v}|^2 -\nabla P + \Gamma_0\nabla^2\mathbf{v}\notag\\ 
&- \Gamma_2\nabla^4\mathbf{v}- (\alpha + \beta |\mathbf{v}|^2)\mathbf{v} - \phi\nabla\mu \ ,
\end{align}
$\Gamma_0$ is the viscosity, and $\Gamma_2$ is a higher-order damping constant. The Toner-Tu model imposes a negative viscosity $\Gamma_0 < 0$ to trigger instabilities, while using $\Gamma_2 > 0$ to stabilize the system at large wave vectors \cite{wensink_meso-scale_2012, dunkel_minimal_2013}. 
The strength of self-advection and alignment between polar flocks is modulated by $S$.
The constants $\alpha$ and $\beta$ represent a cubic velocity potential, where $\text{v} = \sqrt{-\alpha/\beta}$ is the preferred speed of a flock.
We impose the incompressibility condition $\nabla\cdot\mathbf{v} = 0$ to determine the pressure $P$.
The body force from the domains is given by $\phi\nabla\mu$, and facilitates two-way coupling between the Toner-Tu and Cahn-Hilliard equations.
The Toner-Tu model has been used to model biological systems, including gliding actin \cite{Hueschen2022}, chemotactic bacteria \cite{Miller2024}, and qualitatively captures the collective F-actin motion we observe in our experiments [Fig.~\ref{fig:simulations}(a), Supplemental Videos S3-S4].

In this work, we focus on varying $S$ to modulate the strength of the nonlinear term on the left-hand side of Eq.~\ref{eq:TonerTu}, which in turn regulates the strength of convective transport in Eq.~\ref{eq:CahnHilliard}.
We vary $S>0$ to model the puller-type behavior exhibited by gliding actin; large $S$ generates strong convective transport in Eq.~\ref{eq:CahnHilliard} \cite{dunkel_minimal_2013}.
We use pseudo-spectral methods with periodic boundary conditions to numerically evolve Eqs.~\ref{eq:CahnHilliard} and \ref{eq:TonerTu} from initially homogeneous concentration and vorticity fields with small uniform noise.

\begin{figure}
    \includegraphics[width=\linewidth]{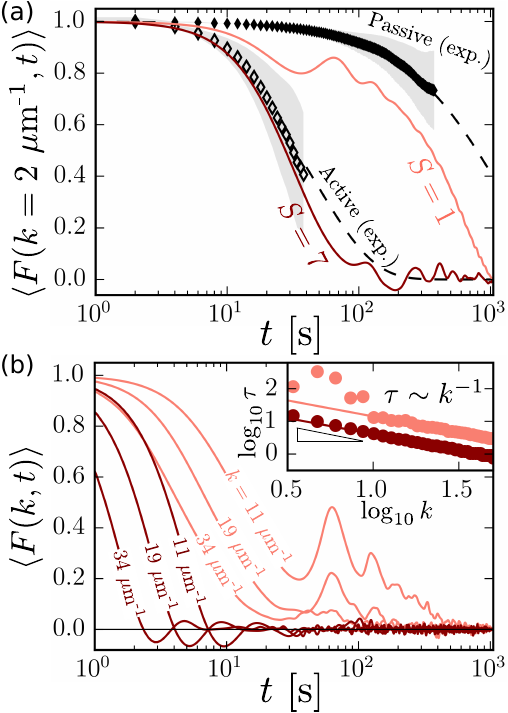}
    \caption{
     Membrane domains coupled to swarming F-actin flows relax $\approx 23 \times$ faster compared to passive membrane domains driven by purely thermal forces.
    (a)
    The average intermediate scattering function $\langle F(k=2\mu\mathrm{m}^{-1}, t) \rangle$ is plotted as a function of lag time $t$ for experiments (black diamonds) and simulations (red curves).
    Experimental data (black diamonds) is plotted for both passive (filled) and active (open) samples.
    $\langle \cdot \rangle$ denotes an average over 30 simulation realizations or 3 experimental replicates.
    Shading represents the standard deviation.
    Dashed lines represent an exponential fit, from which the experimental relaxation time $\tau$ is extrapolated.
    (b)
    Simulated $\langle F(k,t) \rangle$ is plotted as a function of lag time $t$ for various wave vectors $k=$ 11, 19, 34 \textmu m\textsuperscript{-1}. 
    Inset shows simulation relaxation time $\tau$ plotted as a function of wave vector $k$, showing a linear scaling.
    Model parameters for the theory are $(M, \kappa, \Gamma_0, \Gamma_2, \alpha, \beta) = (1, 0.25, -1, 1, -0.2, 1)$. 
    }
    \label{fig:structure_factor}
\end{figure}

\section{Results and Discussion}

To analyze the evolution of domain structure over time, we use differential dynamic microscopy (DDM) \cite{Giavazzi2009,Cerbino2008} to calculate the intermediate scattering function, $F(k,t)$, as a function of wave vector $k$ and lag time $t$ [see Supplemental Material and Supplemental Figs.~S1-S2 for image analysis details].
Fig.~\ref{fig:structure_factor}(a) presents the intermediate scattering function averaged over three experimental realizations for both active (actin and ATP present) and passive (no actin, no ATP [Supplemental Video S2]) bilayers at wave vectors of $\approx$2 \textmu m\textsuperscript{-1}.
We focus on the 2 \textmu m\textsuperscript{-1} wave vector as it emerges to be the dominant relaxation mode across our active samples.

For passive samples, the absence of an underlying actin network frees lipid domains to relax by coarsening and thermal diffusion.
We find that the domains relax slowly, with relaxation time $\tau \approx 1100$ s.
In contrast, in active samples, swarming F-actin couple to the Ld phase, driving rapid surface flows within the fluid bilayer.
This actin-driven advection leads to a $\approx 23 \times$ faster relaxation compared to the passive case, with an average relaxation time $\tau \approx 48$ s for active samples.

The active domains do not relax fully on experimental time scales due to the slowing of F-actin when strongly confined between the bilayer and coverslip.
Additionally, passive domains relax very slowly when larger than $\approx 5-10$ \textmu m in size due to the slow dynamics of $t^{1/3}$ coarsening and the additional drag with the glass.
Lacking access to full relaxation, we use Monte-Carlo simulations to predict the fully relaxed microstructure of a given sample of Lo domains, and thus determine the appropriate magnitude of $F(k,t)$ from DDM data [see Supplemental Material and Supplemental Fig.~S2].
We estimate relaxation time $\tau(k=2$ \textmu m\textsuperscript{-1}) of passive and active samples by fitting the data to an exponential decay $\langle F(k,t) \rangle = A\exp\left[-\alpha(k) t\right]$ [Fig.~\ref{fig:structure_factor}(a), dashed curves], and taking the half life of the fit $\tau(k)=-\ln(1/2)/\alpha(k)$.
Although the analytical form of domain relaxation is not strictly exponential for the active case, we use this form to extract an approximate relaxation time and to quantify the faster dynamics seen in active domains compared to passive domains [Fig.~\ref{fig:structure_factor}(a).]

\begin{figure}
    \includegraphics[width=\linewidth]{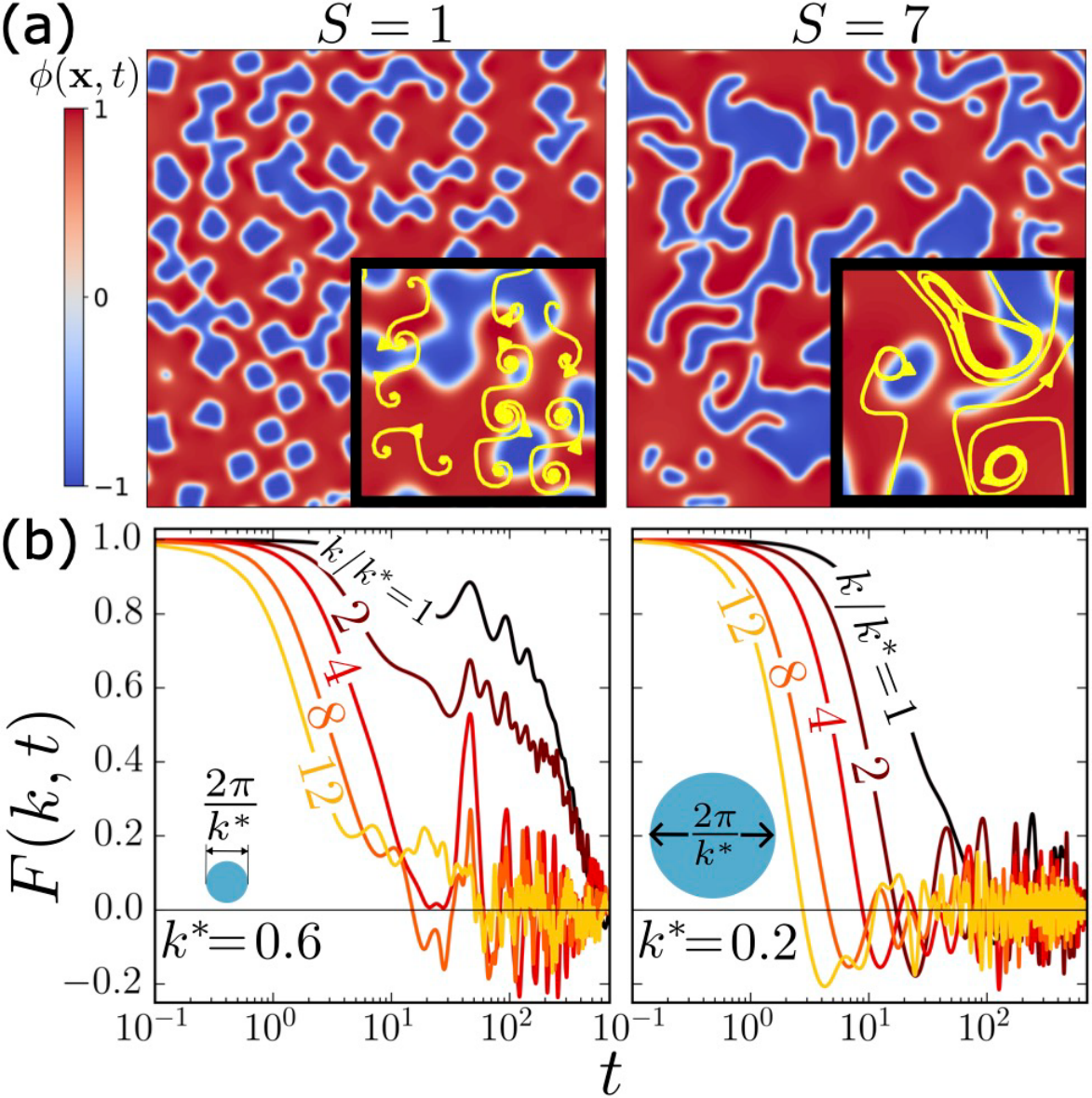}
    \caption{
    Phase domains become entrained in the chaotic eddy patterns of the Toner-Tu flows, generating oscillating waves in the intermediate scattering function.
    (a)
    Numerical solutions of the coupled Toner-Tu---Cahn-Hilliard equations for $S=1$ (left) and $S=7$ (right).
    A snapshot of the phase order parameter $\phi$ is taken at long times, once the system has reached a steady state.
    Insets show velocity streamlines (yellow) from the Toner-Tu flows.  
    (b)
    Intermediate scattering function $F(k, t)$ obtained from a single realization for non-dimensional wave vectors $k/k^*=$ 1, 2, 4, 8, 12. 
    The critical wave vector $k^*$ corresponds to the peak of the static structure factor of the domains [see Supplemental Fig.~S6]. 
    The average size of the domains can be approximated as $2\pi/k^*$.
    Strong convection ($S=7$) results in a $3\times$ increase in domain size compared to weak convection ($S=1$).
    }
    \label{fig:simulations}
\end{figure}

Our numerical results [Fig.~\ref{fig:simulations}(a), Supplemental Videos S3-S4] qualitatively replicate the experimental domain morphology, as actin flows compete with membrane line tension to deform domains into highly non-circular shapes.
Since the Toner-Tu model is phenomenological, our purpose is to qualitatively study the effects of different flow patterns on the phase field evolution.
To compare simulation data with experiments, the physical properties of lipid membranes were fit to the thermodynamic parameters of Eq.~\ref{eq:CahnHilliard}, resulting in a line tension of 0.5 pN, interfacial width of 50 nm, and 2D surface viscosity of 2$\times 10^{-8}$ kg/s [see Supplemental Material for simulation details].
These parameters agree with experimental measurements of lipid membrane properties \cite{cicuta_diffusion_2007,Tian2007,camley_dynamic_2010}.

The $S=7$ curve qualitatively captures the trend we observe in the experimental data for our chosen parameters.
We also numerically evolved Eqs.~\ref{eq:CahnHilliard}-\ref{eq:TonerTu} for $S = 1$, which corresponds to zero self-advection, to compare the $S = 7$ result.
Note that the $S = 1$ case is not meant to model the passive experimental results, which measure structure fluctuations due to Brownian motion.
The solid curves in Fig.~\ref{fig:structure_factor}(a) show that increasing the strength of advection by increasing $S=1$ to $S=7$ accelerates mixing in the phase field for the $k=2$ \textmu m\textsuperscript{-1} wave vector, and thus causes the domains to relax more quickly.
We also derived an expression for the decay of the intermediate scattering function for arbitrary velocity fields, which predicts the non-exponential trend we observe in Fig.~\ref{fig:structure_factor} [see Supplemental Material].

In Fig.~\ref{fig:structure_factor}(b), we show $\langle F(k, t) \rangle$ for different wave vectors and find damped oscillations at long lag times. Fig.~\ref{fig:structure_factor}(b) inset shows that the relaxation time scales as $\tau(k)\sim k^{-1}$ for both weak and strong convection, consistent with a convection-driven domain structure.
The Toner-Tu equations describe chaotic and turbulent-like flows \cite{bratanov_new_2015, wensink_meso-scale_2012},
but despite averaging over 30 simulation realizations, the oscillations in Fig.~\ref{fig:structure_factor}(b) persist.
As with other chaotic systems, two trajectories with nearby initial conditions may rapidly diverge in behavior, and after long times, only statistical statements can be made.
In Fig.~3, we observe that the underlying Toner-Tu velocity field governs the morphology and dynamic structure of the phase field for each individual realization.
For $S=1$, we observe small, fragmented domains that are swept along in tightly-spaced, repeating patterns of eddies  [Fig.~\ref{fig:simulations}(a), yellow streamlines in left inset].
In contrast, for $S=7$ [Fig.~\ref{fig:simulations}(a), right], we observe large, sweeping domains that travel over longer distances before fragmenting and dissipating, commensurate with the longer spacing between the vortices.

We confirm these qualitative observations by finding the magnitude $k^*$ of the wave vector associated with the peak of the static structure factor of the phase and vorticity fields. 
Averaging over 30 realizations, we find that $k^*$ has similar values for both the phase field and vorticity field, increasing $3\times$ from $S=7$ to $S=1$ [Supplemental Fig.~S6].
This increase is consistent with our observation that vortex spacing dictates domain structure and that increasing $S$ coarsens the static structure of vortices and domains.

In Fig.~\ref{fig:simulations}(b), we plot several wave vectors of the intermediate scattering function of the phase field, for a single realization each of $S=1$ and $S=7$ [see Supplemental Fig.~S5 for more realizations].
In a single realization, the trajectories of individual domains have an outsized impact on $F(k,t)$, due to the chaos in the Toner-Tu flows, which becomes apparent in the relatively noisy curves in Fig.~\ref{fig:simulations}(b).

Oscillatory behavior has been observed in the intermediate scattering function for active Brownian particles \cite{dulaney_waves_2020, sevilla_generalized_2021, kurzthaler_intermediate_2016} and bacterial suspensions \cite{zhao_quantitative_2024}.
Here, we hypothesize that the oscillatory features seen in $F(k,t)$ are in part caused by the periodic translation of domains about the repeating vortex positions [Fig.~\ref{fig:simulations}(b) insets], and that the lag time is controlled by $S$.
For $S=7$, $F(k,t)$ reaches zero before the most prominent long-wavelength oscillations appear, suggesting full relaxation [Figs.~\ref{fig:structure_factor}(b) and \ref{fig:simulations}(b)].
We observe that the $S=7$ vortex centers are highly mobile, and that as the eddies translate, they sweep the domains along so that the domains collide and exchange mass freely.
However, in the $S=1$ sample, the vortex centers are more fixed in lattice positions.
As a result, the domains jump around the vortices for several cycles before displacing significantly, and Ostwald ripening plays a larger role in mass exchange between domains [Supplemental Video S3].
Thus, on average, several rotations occur before a domain translates, and the intermediate scattering function oscillates several times before reaching zero.

\begin{figure}
    \centering
    \includegraphics[width=\linewidth]{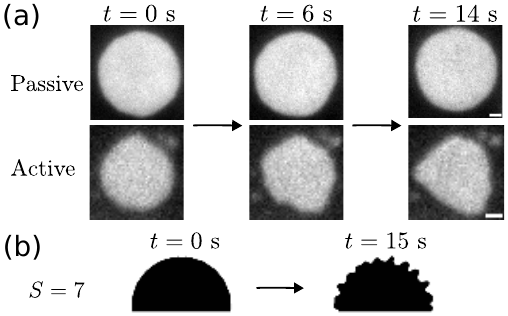}
    \caption{
    Interfacial deformations are shown with and without the presence of activity.
    (a)
    Experiment snapshots of fluctuating passive (top row) and active (bottom row) domains at times $t=0,\ 6,\ 14$ s.
    Scale bar is 2 \textmu m.
    (b)
    Numerical solution snapshots for active domains at $t=0,\ 15$ s for $S=7$. 
    Initially-circular domains are cropped to improve clarity of fluctuations.
    The model parameters used here are: $(M, \kappa, \Gamma_0, \Gamma_2, \alpha, \beta, S) = (1, 0.25, -1, 1, -0.2, 1, 7)$.
    }
    \label{fig:heightFluctuation}
\end{figure}

We have thus far focused on the distribution of an ensemble of membrane domains via the intermediate scattering function. 
However, $F(k, t)$ also measures the density fluctuations at the level of a single domain interface for large $k/k^*$ values.
To better understand the effects of active swarming flows on the evolution of an individual domain interface, we measured the capillary height fluctuation spectra of domains in our experiments and theoretical model [Fig.~\ref{fig:heightFluctuation}, see Supplemental Fig.~S8 for analysis details].

In our experiments, we find that for initially unperturbed domains, interfacial fluctuations of wave vector $k=2$ \textmu m\textsuperscript{-1} grow $\approx$2-4$\times$ over a period of ten seconds after activity is triggered [Fig.~\ref{fig:heightFluctuation}(a), Supplemental Fig.~S9, Supplemental Video~S6].
In some cases, these domain fluctuations lead to fission or pinch-off events [Supplemental Video~S7], while in others, weaker or more transient actin-domain interactions cause heights to plateau and/or fluctuate about some non-circular state.
Meanwhile in passive domains, we find thermal fluctuations to be rapidly dampened by line tension [Fig.~\ref{fig:heightFluctuation}(a), Supplemental Fig.~S9, Supplemental Video~S8].

In our theoretical model, we numerically evolve a single domain using Eqs.~\ref{eq:CahnHilliard} and \ref{eq:TonerTu} to capture the height fluctuation spectrum [see Supplemental Material for height calculation details].
We observe that the Toner-Tu flows create protrusions at the interface, characteristic of capillary instabilities, as seen in mixtures of active and passive fluids \cite{Xu2023,Adkins2022,Caballero2022,Tayar2023}.
These protrusions grow and undulate along the interface before pinching off to create highly deformed structures, consistent with experimental observations [Fig.~\ref{fig:heightFluctuation}(b), Supplemental Fig.~S9, Supplemental Video~S5].
We see these observations reflected in $F(k, t)$ curves in Fig.~\ref{fig:simulations}(b).
As we increase $k/k^*$, the frequency of oscillations increases, which corresponds to the small fragmented domains that are swept in the repeating eddies of the Toner-Tu flows. 
We observed that activity drives membrane relaxation by both advecting whole domains and driving capillary instabilities within existing domains.

\section{Conclusions}
We present an active 2D multiphase fluid whose domains relax $23\times$ faster when advected by flocking actin flows, than under thermal Brownian motion.
We qualitatively recapitulate the swarming behavior of actin using the Toner-Tu equation, and achieve quantitative agreement when increasing the strength of convection ($S$) to reduce the domain relaxation time.
In individual simulations, we observe oscillations in domain structure that decay more quickly as $S$ is increased, consistent with greater turbulence rapidly displacing actin vortices.
Finally, we show that actin flows drive capillary fluctuations at the boundary of the domains, which sometimes pinch off and fragment into smaller domains.

We demonstrate that flocking actin flows can actively mix phase-separated lipid membrane domains, both accelerating structural relaxation and deforming the domains away from their ground-state circular shape.
We complement earlier work that used a contractile network of actin and myosin II to drive continuous growth \cite{ArnoldGubbala2023} by mixing domains with actin flows that do not appreciably change their size.
Given both the role the actin cytoskeleton plays in organizing the plasma membrane \cite{Kusumi2012,Viola2007}, and the importance of Toner-Tu flocking actin flows in controlling motility of certain cells \cite{Hueschen2022}, our results may describe a potential mechanism by which cells actively mix plasma membrane domains.

Our results are also significant in that they represent the first experimental realization of active polar flocks in contact with a 2D multiphase fluid.
Some prior 3D experiments have had to chemically cross-link active filaments in the continuous phase to the dispersed lipid droplets, thus augmenting the relatively weak transmission of stress through continuous phase flows \cite{Tayar2023}.
Although we cannot directly compare our experiments to such 3D systems, we note that in 2D materials, flows and defects decay over much longer distances than in 3D.
Thus, 2D fluids like ours may be uniquely effective at mechanically transmitting stress between active filaments and fluid droplets, due to the loss of a third spatial degree of freedom for lipid relaxation.

\begin{acknowledgments}
This material is based upon work supported by the National Science Foundation under Grant No.~2150686.
This work was supported by the MRSEC Program of the National Science Foundation under Award No. DMR 2308708.
D.P.A. is supported by the National Science Foundation Graduate Research Fellowship under Grant No.~2139319.
S.C.T. is supported by the Packard Fellowship in Science and Engineering.
\end{acknowledgments}

\section*{Author contributions}
A.G. and D.P.A. contributed equally to this work.
A.G., D.P.A., and S.C.T. conceptualized the work;
D.P.A., A.J., and S.A. performed experiments;
D.P.A. analyzed experimental data;
A.G. and S.C.T. developed theory;
A.G. performed numerical simulations;
S.C.T. supervised the study;
and A.G., D.P.A., and S.C.T. wrote the paper.

\end{document}